\documentclass[twocolumn,aps,amsmath,amssymb,color,superscriptaddress,prl,footinbib]{revtex4-1}
\usepackage{stix}
\usepackage{graphicx}
\usepackage{dcolumn}
\usepackage{bm}
\usepackage{times}
\usepackage{tabularx}
\usepackage{footmisc}
\usepackage{xcolor}

\newcommand{\red}[1]{{\leavevmode\color{black}#1}}

\usepackage[normalem]{ulem}

\newcommand{\mb}[1]{\mathbf{#1}}
\newcommand{\mc}[1]{\mathcal{#1}}
\newcommand{\average}[1]{\langle #1 \rangle}

\begin{document}

\title{Hidden Physical Effects in Non-centrosymmetric Crystals}

\author{Zuzhang Lin}
\affiliation{Institute for Advanced Study, Tsinghua University, Beijing 100084, China}
\affiliation{State Key Laboratory of Low Dimensional Quantum Physics and Department of Physics, Tsinghua University, Beijing, 100084, China}

\author{Chong Wang}
\email{ch-wang@outlook.com}
\affiliation{Institute for Advanced Study, Tsinghua University, Beijing 100084, China}
\affiliation{State Key Laboratory of Low Dimensional Quantum Physics and Department of Physics, Tsinghua University, Beijing, 100084, China}
\affiliation{Department of Physics, Carnegie Mellon University, Pittsburgh, Pennsylvania 15213, USA}

\author{Yong Xu}
\email{yongxu@mail.tsinghua.edu.cn}
\affiliation{State Key Laboratory of Low Dimensional Quantum Physics and Department of Physics, Tsinghua University, Beijing, 100084, China}
\affiliation{Frontier Science Center for Quantum Information, Beijing 100084, China}
\affiliation{RIKEN Center for Emergent Matter Science (CEMS), Wako, Saitama 351-0198, Japan}

\author{Wenhui Duan}
\affiliation{Institute for Advanced Study, Tsinghua University, Beijing 100084, China}
\affiliation{State Key Laboratory of Low Dimensional Quantum Physics and Department of Physics, Tsinghua University, Beijing, 100084, China}
\affiliation{Frontier Science Center for Quantum Information, Beijing 100084, China}

\date{\today}

\begin{abstract}
Symmetry forbidden effects in crystals may emerge in a local environment that breaks the symmetries. Yet these hidden physical effects were only discussed in centrosymmetric crystals. Here we propose that hidden physical effects can be generalized to almost all crystallographic symmetric groups and hence reveal their universality and diversity. We systematically discuss certain symmetries in crystals that may induce hidden spin polarization (HSP), hidden berry curvature and hidden valley polarization, with a focus on a specific pattern of HSP whose winding directions of the in-plane spin vectors are the same for adjacent bands, dubbed anomalous HSP. Such an unprecedented spin pattern  arises from relatively weak spin-dependent inter-sector interaction and is demonstrated in mirror-symmetric InSe by first-principles calculations. Unique electric field-dependent splitting into specific spatial polarization pattern may serve as an experimental signature of anomalous HSP. Our results reveal abundant hidden physical effects beyond centrosymmetric crystals and provide new platforms to discuss them for emergent physical effects and future applications.
\end{abstract}

\maketitle

With time reversal symmetry and inversion symmetry, all electronic energy bands are doubly degenerate in crystalline solids. In comparison with spin polarized materials, these materials exhibit much more subtle effects of spin-orbit coupling (SOC). As such, these materials are largely ignored in the studies of SOC induced phenomena, including the Dresselhaus effect \cite{Dresselhaus}, Rashba effect \cite{Bychkov}, anomalous Hall effect \cite{xiao2010berry} and valley polarization \cite{zeng2012valley, mak2012control}. However, recent research efforts have been put into uncovering hidden physical effects induced by the local asymmetry in the centrosymmetric systems. One important advancement is the discovery of hidden spin polarization (HSP), including the hidden Rashba effect and the hidden Dresselhaus effect \cite{zhang2014hidden,Yuan,slawinska2016hidden, zhang2019symmetry}. This discovery has soon inspired studies on other hidden physical effects, such as hidden orbital polarization \cite{Ryoo}, hidden valley polarization (HVP) \cite{liu2015intrinsic} and hidden Berry curvature (HBC)\cite{Cho, schuler2020local}.

Among hidden physical effects, HSP shows a lot of advantages. Unlike the conventional Rashba effect, hidden Rashba effect can be easily manipulated by an external electric field \cite{yao2017direct}, and is proposed to be the key ingredient in the next generation of spin-field effect transistors \cite{Wu,liu2013tunable}. In addition, HSP is believed to bring exotic physical insights into topological insulators \cite{Das}, topological superconductivity \cite{Nakosai} and may even into high-temperature superconductivity \cite{Gotlieb}. Experimentally, HSP has been observed via various methods \cite{Riley,Razzoli,Bawden}. Despite these recent advances, discussion of hidden physical effects has been limited to centrosymmetric crystals. This restriction can be lifted by considering other crystallographic symmetries. For example, because spin is a pseudovector, the mirror symmetry forbids in-plane spin component in mirror-symmetric planes of the Brillouin zone (BZ). However, by projecting the Bloch wave functions onto a mirror-asymmetric sector of the crystal, hidden in-plane spin component may emerge.

\begin{figure}[t]
\centering
\includegraphics[width=1\columnwidth]{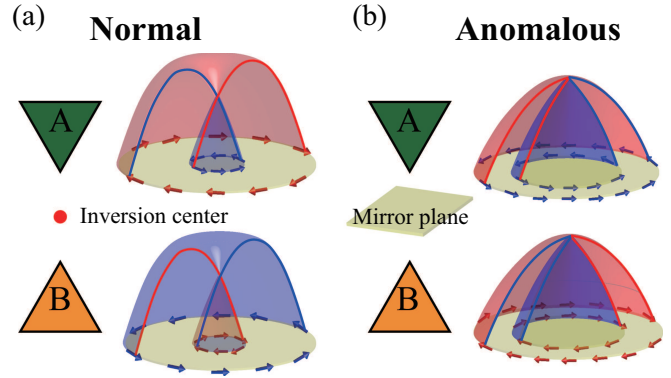}
\caption{Schematic diagrams of (a) normal HSP and (b) anomalous HSP, where sectors A and B of the system are, respectively, related by an inversion center and a mirror plane. The winding direction of spin vectors of two spin split bands are opposite for normal HSP but same for anomalos HSP.}\label{fig:HSP_sketch}
\end{figure}
In this work, we systematically investigate HSP, HBC and HVP under certain crystalline symmetries, including inversion ($I$), mirror ($m$), two-fold rotation ($C_2$), rotation-reflection (\red{$S_n$ }in Schoenflies symmetry notation \cite{dresselhaus2007group}) \red{and their combination with time reversal symmetry $T$}. We find hidden physical effects are allowed in most point groups and candidate materials can be easily discovered with the guidance of symmetry analyses. As an example, we demonstrate the HSP in a mirror-symmetric crystal InSe by first-principles calculations. Unexpectedly, the winding directions of the in-plane spin vectors are the same for adjacent bands, as opposed to the Rashba effect, where the spin-split bands have opposite winding directions. We develop a $\mb{k \cdot p}$ model and attribute the anomalous HSP to the relatively weak SOC  between the two sectors permuted by mirror symmetry. We also predict the experimental signatures of anomalous HSP. These findings enrich our understanding of hidden physical effects and uncover a large amount of candidate materials for experimental observation.

\emph{Symmetries that allow hidden physical effects.}---When a physical quantity is forbidden by crystallographic space time symmetry, nonvanishing effects may be found in a local environment that breaks this symmetry. This observation is important since many experimental detection techniques measure local quantities, effectively breaking the global symmetry that enforces the stringent constraint on this physical quantity. \red{We start our discussion with a general spacetime symmetry $\mc{S}$, which keeps some crystal wave vectors $\mb{k}$ invariant. These wave vectors constitute a manifold which we will denote by $\mc{M}[\mc{S}]$. A general observable $\mc{O}$ measured by energy and momentum resolved experimental techniques (such as angle-resolved photoemission spectroscopy) takes the following form}
\begin{equation}
    \red{\average{\mc{O}}_{\mb{k}} = \sum_{\psi_{\mb{k}}} \langle \psi_{\mb{k}} |\mc{O}| \psi_{\mb{k}} \rangle,\label{observable}}
\end{equation}
\red{where $\psi_{\mb{k}}$ is summed over degenerate states within a band. If the band is not degenerate, the summation is not necessary. We will focus on one band at a time such that we will not write out the band index explicitly.}

\red{$\average{\mc{O}}_{\mb{k}}$ is severely constrained by $\mathcal{S}$ on $\mathcal{M}[\mathcal{S}]$. For example, since space-time inversion symmetry $(IT)$ flips spin $\mb{s}$, for $IT$-symmetric crystals, $\average{\mb{s}}_{\mb{k}}=\mb{0}$ for any $\mb{k}$ ($\mc{M}[IT]$ is the whole Brillouin zone). However, if we divide the space into two regions A and B, such that A and B are permuted by $IT$, the following relation holds}
\begin{equation}
\average{\mb{s}^A}_{\mb{k}} = -\average{\mb{s}^B}_{\mb{k}}.\label{normalHSP}
\end{equation}
\red{Here, $\mb{s}^{A(B)} = P^{A(B)} \mb{s} P^{A(B)}$ and $P^{A(B)}$ is the projection operator projecting to the A(B) region. Equation (\ref{normalHSP}) shows that although the total spin polarization is forbidden by $IT$, the spin polarization may be detected locally, taking opposite values for regions A and B.} This phenomenon, proposed by Ref. [\onlinecite{zhang2014hidden}], is illustrated in Fig.~\ref{fig:HSP_sketch}(a). \red{Hereafter, we will refer to A and B as sectors}.


For a tensor $\mc{T}$ with rank larger than or equal to $1$, the above relation may be generalized to explicitly include specific Cartesian components. \red{For a specific Cartesian index $i$, if $\mc{T}_{i}$ is flipped by $\mc{S}$, $\average{\mc{T}_i}_{\mb{k}}=0$ on $\mc{M}[\mc{S}]$ for $\mc{S}$-symmetric crystals. However, if sectors A and B are permuted by $\mc{S}$, the constraint due to $\mc{S}$ reads}
\begin{equation}
  \red{  \average{\mc{T}_i^A}_{\mb{k}} = - \average{\mc{T}_i^B}_{\mb{k}},\label{generalized-hidden}}
\end{equation}
\red{indicating nonvanishing $\mc{T}_i$ localized on A and B sectors. Equation (3) indicates a plethora of hidden physical effects in non-centrosymmetric crystals, including HSP, HBC and HVP, which have been overlooked in previous research efforts. In the following, we restrict our discussion to symmetries that permute two sectors. Important examples include $m$, $m T$, $C_2$, $C_2 T$ and $S_n T$ ($\red{n=3,4,6}$).}


\begin{figure}[t]
\centering
\includegraphics[width=1\columnwidth]{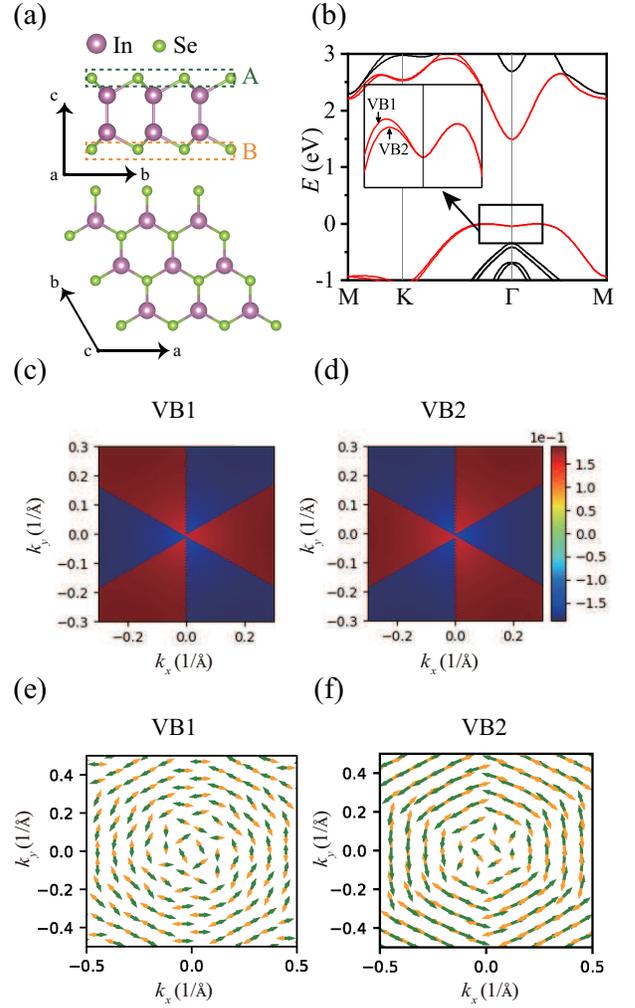}
\caption{Crystal structure, band structure and spin polarization of monolayer InSe. (a) Sideview and topview of the crystal structure. (b) Band structure of monolayer InSe and the inset is magnification of the details of the two highest valence bands. The out-of-plane spin polarization of VB1 (c) and VB2 (d), which is represented by color scale. The in-plane spin polarization projected on sector A (green color) and  sector B (orange color) for VB1 (e) and VB2 (f).}
\end{figure}
As an example, in Eq. (\ref{generalized-hidden}), if \red{$\mc{S}$} is chosen as mirror symmetry in the $z$ direction ($m_z$), spin polarizations \red{$\average{s_x}_{\mb{k}}$} and \red{$\average{s_y}_{\mb{k}}$} will be forbidden in the $m_z$-invariant $k_z=0$ plane. If \red{$\mc{S}$} is chosen as \red{$m_z T$}, \red{$\average{s_z}_{\mb{k}}$} will be forbidden in the \red{$m_z T$}-invariant $k_x=k_y=0$ line. The above statements clarify the possibility of HSP in mirror-symmetric materials. Analogously, crystals with $C_2$, \red{ $C_2T$} or $S_n T$ symmetries can also exhibit HSP in certain lines or planes in the reciprocal space.

Similar symmetry analyses can also be applied to HBC and HVP (see Note S1 in the Supplemental Material \footnote{\label{note1}See Supplemental Material at [URL will be inserted by publisher] for other HSP patterns, discussions on HBC and HVP, calculation methods, symmetry analyses of the Hamiltonian and discussion on inter-sector SOC, which includes Refs. \cite{xiao2010berry, yao2008valley, kresse1993ab, kresse1996efficiency, kresse1999ultrasoft, perdew1996generalized}}).  With the inclusion of these symmetries, hidden physical effects may be found in crystals with almost all point groups, which we summarize in Table S1 in the Supplemental Material \cite{Note1}. Under these guidances, we find HSP in monolayer InSe, monolayer In$_2$Se$_3$ and bulk material GeSe$_2$, which are examples of $m_z$-, $C_{2}$- and $S_{4z} T$-symmetric crystals, respectively. In monolayer InSe, hidden in-plane spin polarization with unconventional spin textures can be found, as discussed in detail in the following section. Monolayer In$_2$Se$_3$ and bulk GeSe$_2$ show HSP along certain high symmetry lines in BZ, the discussion of which is deferred to Fig. S1 and Fig. S2 in the Supplemental Material, respectively.

\begin{figure*}[t]
\centering
\includegraphics[width=1\textwidth]{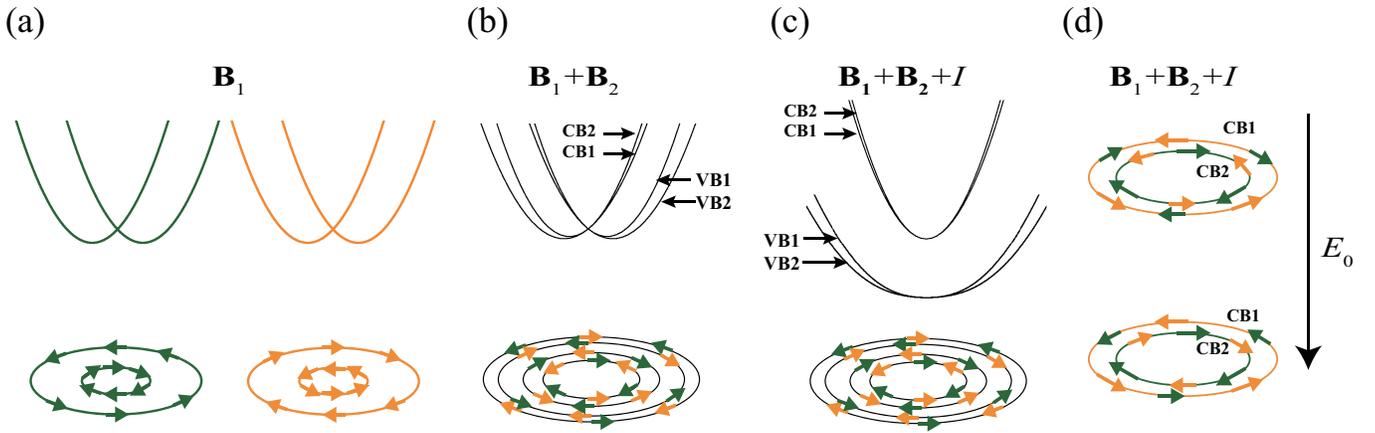}
\caption{Schematic diagrams of band splittings and corresponding spin polarization in the mirror-symmetric crystals. (a) The two sets of Rashba bands (degenerate in energy) located on sector A (green) and sector B (orange) with Rashba-type spin textures when only the intra-sector SOC  $\mb{ B_{1}}(\mb{ k})$ is included.  (b) The four splitting bands (upper panel) and the spin textures (lower panel) when the weak inter-sector SOC $\mb{ B_{2}}(\mb{ k})$) is also included. (c) The bands CB1, CB2, VB1 and VB2 (upper panel) with spin textures (lower panel) when spin-independent inter-sector-interaction $I(\mb{ k})$ is further included. (d) The spin textures of bands CB1 and CB2 when the zero-order $E_0$ term of \red{Eq. (5)} in the main text is further included. The arrow in (d) indicates $E_0$ increases from the upper panel to the lower panel.  The green and orange curves in (a) and (d)  refer to the bands  mainly contributed by sector A and  B, respectively, while the black curves in (b) and (c) refer to the bands  equally contributed by sector A and  B.  The green arrows and orange arrows represent the in-plane spin polarization contributed by sector A and sector B, respectively.}
\end{figure*}

\emph{HSP in monolayer InSe.}---In the following, we study monolayer InSe as an example of HSP protected by mirror symmetry. Monolayer InSe (space group $P \overline{6} m 2$) has a mirror plane perpendicular to the $z$ direction (we set the $c$ axis along the $z$ direction) and the mirror plane permutes the top Se atoms and the bottom Se atoms, which we will refer to as sectors A and B [Fig. 2(a)]. Monolayer InSe is a semiconductor with an indirect band gap of 1.49 eV [Fig. 2(b)]. Its highest valence band along the $\Gamma$-K path splits into two bands (VB1, VB2) due to SOC. The maximum amplitude of this splitting is about 10 meV. By first-principles calculations (see Note S2 in the Supplemental Material \cite{Note1}), we find out-of-plane spin polarization [Figs. 2(c) and 2(d)] for each band. Due to $m_z$ symmetry, the total in-plane spin polarization of each band is zero, but HSP can be revealed by projecting each band onto sector A or sector B [Figs. 2(e) and 2(f)], as discussed in the previous paragraphs.


Beyond our expectation, the orientations of spin textures projected onto sector A (or B) of VB1 and VB2 are both anticlockwise (clockwise) [Figs. 2(e) and 2(f)], realizing a sector-spin locking. Similarly, the spin textures of the two conduction bands (CB1 and CB2) are both clockwise and anticlockwise when projected onto sector A and sector B, respectively (see Fig. S3 in the Supplemental Material \cite{Note1}). Similar HSP is observed in InSe-structure like materials GaS and GaSe (see Fig. S4 and Fig. S5 in the Supplemental Material \cite{Note1}). Since the splitting between the two valence bands of InSe stems from SOC, it is quite unusual to get the same orientation of the spin textures of these two bands. Therefore, we refer to it as the anomalous HSP, while the HSP in the centrosymmetric system found in Ref. [\onlinecite{zhang2014hidden}] is dubbed as normal. To compare the normal and anomalous HSP clearly, we illustrate them in Fig. 1.

\emph{ $\mb{k} \cdot \mb{p}$ model for anomalous HSP.}---To understand the anomalous HSP discussed above in monolayer InSe, a $\mb{ k} \cdot \mb{p}$ Hamiltonian around $\Gamma$ point is derived based on the symmetry analyses. The basis functions are $|A,\uparrow\rangle$, $|A,\downarrow\rangle$, $|B,\uparrow\rangle$ and $|B,\downarrow\rangle$, where $A$ and $B$ refer to sectors A and B, and spin up (down) state is described by $\uparrow$ ($\downarrow$). The $\mb{k} \cdot \mb{p}$ Hamiltonian invariant under all these symmetric transformations is  (see Note S3 in the Supplemental Material \cite{Note1} for more details)
\begin{equation}
H(\mb{k})=c(\mb{k}) \tau_{0} \sigma_{0}+\mb{ B}_{1}(\mb{k}) \cdot \bm{\sigma}+\mb{B}_{2}(\mb{k}) \cdot \bm{\sigma}+I(\mb{k}) \tau_{x} \sigma_{0},
\end{equation}
where the Pauli matrices $\tau$ and $\sigma$ describe the sector and spin degrees of freedom, respectively. we define $\bm{\sigma}=(\sigma_x,\sigma_y,\sigma_z)$. The energy dispersion  $c(\mb{k})$ term  and spin-independent inter-sector-interaction $I(\mb{k})$ term describe the Hamiltonian without SOC. Here, $\mb{ B_{1}}(\mb{k})$ and $\mb{ B_{2}}(\mb{k})$ are two SOC-induced effective magnetic fields, with $\mb{ B_{1}}(\mb{k})=R \tau_{z}\left(-k_y, k_{x},0\right)$  and $ \mb{ B_{2}}(\mb{k})=M \tau_{y}\left(k_x^{2}-k_{y}^{2},-2 k_{x} k_{y},0\right)$. $\mb{ B_{1}}(\mb{k})$ and $\mb{ B_{2}}(\mb{k})$ arise from the intra-sector and inter-sector SOC, respectively. All coeffcients in the Hamiltonian are real.

Note that $c(\mb{k})$ describes the average dispersion of the four relevant bands. With intra-sector SOC [$\mb{ B_{1}}(\mb{k})$], we get two sets of doublely degenerate Rashba split bands separately located on sector A and sector B. This is depicted by the upper panel and lower panel in Fig. 3(a), respectively. Since the $\mb{ B_{1}}(\mb{k})$ term respects inversion symmetry, the effective magnetic fields $\mb{ B_{1}}(\mb{k})$ of two sectors compensate each other and the total in-plane spin polarization is zero.

When the inter-sector SOC  is turned on [$\mb{ B_{2}}(\mb{k})$], inversion symmetry is broken and the degeneracy of the two sets of Rashba bands will be lifted [Fig. 3(b)]. Each doubly degenerate band splits into two descendant non-degenerate bands, leading to VB2, VB1, CB1 and CB2 as labeled in Fig. 3(b). Each of these four bands is contributed by both sectors A and B. The two descendant bands VB2 and VB1 inherit the HSP from their parent band so that they possess the same winding orientation. With spin-independent inter-sector interaction is further included [$I(\mb{ k})$ in $H(\mb{ k})$ ], the persistent degeneracy at $\Gamma$ is also lifted and a large band splitting may arise as long as the $I(\mb{ k})$ term is large enough [Fig. 3(c)].

If the $\mb{ B_{2}}(\mb{k})$-related band splitting is relatively smaller compared to the band splitting caused by intra-sector SOC interaction [described by $\mb{ B_{1}}(\mb{k})$], the orientations of HSPs of two inter-sector SOC-split subbands are the same as shown in Fig. 3(c), which explains the same orientations of HSPs of two SOC-split subbands VB1 and VB2 in InSe. \red{Such anomalous HSP exhibits experimentally detectable unique spin patterns under an external electric field, as discussed in the following section.}

\emph{Electric control of HSP.}---The mirror symmetry of the system can be further broken under an external electric field along the $z$ direction, leading to another Hamiltonian $H_{E}(\mb{ k})$ (up to the first order of $k$):
\begin{equation}
\begin{array}{r}
H_{E}(\mb{ k})=E_0\tau_{z} \sigma_{0}+\mb{ B}^{{\rm ext}}_{1}(\mb{ k}) \cdot \bm{\sigma}+\mb{ B}^{{\rm ext}}_{2}(\mb{ k}) \cdot \bm{\sigma},
\end{array}
\end{equation}
where $\mb{ B}^{{\rm ext}}_{1}(\mb{ k})$ and $\mb{ B}^{{\rm ext}}_{2}(\mb{ k})$ are another two effective magnetic fields induced by the external electric field, with $ \mb{ B}^{{\rm ext}}_{1}(\mb{ k})=E_1 \tau_{0}(-k_y,k_x,0 )$ and  $\mb{ B}^{{\rm ext}}_{2}(\mb{ k})=E_2\tau_{x}(-k_y,k_x,0)$. All parameters are real.

The usually dominated zero-order $E_0$ term determines the energy difference between the two sectors because of the external electric field, resulting in sector polarization for each band \cite{gong2013magnetoelectric,du2019strongly}. In the following, we show the evolution of the patterns of the spin textures of the two conduction bands under an electric field in the positive $z$ direction. When the electric field is small, the spin polarizations projected on sectors A and B remain almost the same but each band now has net spin polarization because it is dominantly contributed by either sector A or sector B, as illustrated in the upper panel in Fig. 3(d). A larger electric field will induce normal Rashba effect. Therefore, after a critical field, the winding directions of the in-plane spin vectors of CB1 and CB2 become the same for two sectors [the lower panel in Fig. 3(d)], which are verified through first-principles calculations  (see Fig. S6  in the Supplemental Material \cite{Note1}). The calculated spin polarization under external electric fields through $\mb{ k \cdot p}$ model reproduces very well the results of first-principles calculations (see Fig. S7 in the Supplemental Material \cite{Note1}).  The evolution of the spin polarizations of the two valence bands under an electric field should be analogous. The evolution of the spin polarizations of the four bands can be used as a signature for experimental detection of the anomalous HSP. The above analysis has assumed small $\mb{ B}^{{\rm ext}}_{1}(\mb{ k})$ and $\mb{ B}^{{\rm ext}}_{2}(\mb{ k})$. In the case of large $\mb{ B}^{{\rm ext}}_{1}(\mb{ k})$ and $\mb{ B}^{{\rm ext}}_{2}(\mb{ k})$, the spin and sector polarization patterns might be different, the discussion of which is presented in Fig. S8 and Fig. S9 of the Supplemental Material \cite{Note1}.

In summary, we have revealed the presence of hidden physical effects in the \red{$m$-, $mT$-, $C_{2}$-, $C_{2}T$  and  $S_nT$- symmetric crystals}, and thus uncovered the abundance of hidden physical effects' candidate materials whose point groups contain these symmetry operators. Based on first-principles calculations, we carefully study HSP in mirror-symmetric InSe and predict the emergence of anomalous HSP. The possibility of anomalous HSP in other symmetry and anomalous phenomena of other hidden physical effects remains an open question. We also predict unique tunable patterns of the spin textures of such systems under an external electric field, which can be used for  experimental detection. Besides the electric field, other external field such as magnetic field and strain should also be able to tune the spin polarization. Our results enrich the physical insights of hidden physical effects and predict the existence of a large amount of candidates, which may bring exotic physical phenomena and benefit further applications in spintronics.

This work was supported by the Ministry of Science and Technology of China (Grant Nos. 2016YFA0301001, 2018YFA0307100 and 2018YFA0305603), the National Natural Science Foundation of China (Grant Nos. 11674188, 51788104 and 11874035), and the Beijing Advanced Innovation Center for Future Chip (ICFC).

\end{document}